\documentclass[prl,twocolumn,superscriptaddress,amsfonts,amssymb,showpacs,a4]{revtex4}

\usepackage{graphicx}%
\usepackage{dcolumn}
\usepackage{amsmath}
\usepackage{dcolumn}% Align table columns on decimal point
\usepackage{hyperref}
\usepackage{times}
\usepackage{xcolor}
\usepackage{soul}
\usepackage{bm}

\usepackage{draftcopy}
\draftcopySetGrey{0.7}

\makeatletter
\def\btt#1{\texttt{\@backslashchar#1}}%
\DeclareRobustCommand\bblash{\btt{\@backslashchar}}%
\makeatother

\begin{document}
\sethlcolor{yellow}

\title{CeFePO: \textit{f}-\textit{d} hybridization and quenching of superconductivity}

\author{M.G. Holder}
\affiliation{Institut f\"ur Festk\"orperphysik, Technische Universit\"at Dresden, D-01062 Dresden, Germany}

\author{A. Jesche}
\affiliation{Max-Planck-Institut f\"ur Chemische Physik fester Stoffe, D-01187 Dresden, Germany}

\author{P. Lombardo}
\author{R. Hayn}
\affiliation{Laboratoire Mat\'eriaux et Micro\'electronique de Provence associ\'e au Centre National de la Recherche Scientifique, 13397 Marseille, France}

\author{D. V. Vyalikh}
\author{S. Danzenb\"acher}
\author{K. Kummer}
\affiliation{Institut f\"ur Festk\"orperphysik, Technische Universit\"at Dresden, D-01062 Dresden, Germany}

\author{C. Krellner}
\affiliation{Max-Planck-Institut f\"ur Chemische Physik fester Stoffe, D-01187 Dresden, Germany}

\author{C. Geibel}
\affiliation{Max-Planck-Institut f\"ur Chemische Physik fester Stoffe, D-01187 Dresden, Germany}

\author{Yu. Kucherenko}
\affiliation{Institute for Metal Physics, National Academy of Sciences of Ukraine, UA-03142 Kiev, Ukraine}

\author{T. Kim}
\affiliation{Institut f\"ur Festk\"orperforschung, IFW Dresden, D-01171 Dresden, Germany}

\author{R. Follath}
\affiliation{Helmholtz-Zentrum Berlin f\"ur Materialien und
Energie GmbH, Elektronenspeicherring BESSY II, D-12489 Berlin,
Germany}

\author{S. L. Molodtsov}
\affiliation{Institut f\"ur Festk\"orperphysik, Technische Universit\"at Dresden, D-01062 Dresden, Germany}

\author{C. Laubschat}
\affiliation{Institut f\"ur Festk\"orperphysik, Technische Universit\"at Dresden, D-01062 Dresden, Germany}

\date{\today}

\begin{abstract}
Being homologue to the new, Fe-based type of high-temperature
superconductors, CeFePO exhibits magnetism, Kondo and
heavy-fermion phenomena. We experimentally studied the electronic
structure of CeFePO by means of angle-resolved photoemission
spectroscopy. In particular, contributions of the Ce $4f$-derived
states and their hybridization to the Fe $3d$ bands were explored
using both symmetry selection rules for excitation and their
photoionization cross-section variations as a function of photon
energy. It was experimentally found $-$ and later on confirmed by
LDA as well as DMFT calculations $-$ that the Ce~4$f$ states
hybridize to the Fe 3$d$ states of $d_{3z^2-r^2}$ symmetry near
the Fermi level that discloses their participation in the
occurring electron-correlation phenomena and provides insight into
mechanism of superconductivity in oxopnictides.
\end{abstract}


\maketitle

The unusual superconducting properties of the novel Fe-based
oxopnictides with transition temperatures ($T_c$) up to 55\,K have
attracted considerable attention
\cite{Kamihara08,Chen08,Ren08,GFChen08}. While pure $R$FeAsO ($R$:
rare-earth elements) compounds reveal metallic properties, doping
by F on O sites leads to superconductivity. The proximity of the
superconducting state to spin-density wave formation gave rise to
speculations that the underlying pairing mechanism is based on
magnetic fluctuations \cite{Mazin2009}. Superconductivity without
doping, although at reduced $T_c$ with respect to the
arsenides, is found in the isoelectronic phosphides, except for
$R$=Ce~\cite{Kamihara2008, Baumbach2009}. In CeFeAsO both Fe and
Ce order antiferromagnetically below a Neel temperature of
140\,K~\cite{Zhao2008} and 3.7\,K~\cite{Jesche2009}, respectively.
A gradual replacement of As by P leads first to the vanishing of
the Fe magnetism, coupled with a change of the Ce order to
ferromagnetism \cite{Luo2009}. For further P doping the Ce order
is suppressed, resulting in a paramagnetic heavy-fermion compound
\cite{Bruning2009}.

This wide variation of properties is a consequence of a strong
sensitivity of the valence-band (VB) structure to the lattice
parameters and to interaction with localized $f$ states. Close to
the Fermi level ($E_F$) the electronic structure of $R$Fe$Pn$O
($Pn$: phosphorus or arsenic) materials is dominated by five
energy bands that have predominantly Fe $3d$ character
\cite{Vildosola2008,Kuroki2009}. Small variations of the lattice
parameters affect particularly two of these bands, namely those
containing $d_{xy}$ and $d_{3z^2-r^2}$ orbitals. Increasing the
distance of the pnictogen ions to the Fe plane shifts the
$d_{xy}$-derived band towards lower and the $d_{3z^2-r^2}$-derived
bands towards higher binding energies (BE) leading to a transition
from 3D to 2D behavior of the Fermi surface (FS). As discussed in
Ref.~[\onlinecite{Kuroki2009}], superconductivity delicately
depends on nesting conditions between the FS sheets generated by
the above mentioned bands around the $\Gamma$ point and those
located around the $M$ point in the Brillouin zone (BZ). The
nesting conditions may be affected by variations of the lattice
parameters or interaction with 4$f$ states.

Purpose of the present work is to study the electronic structure
of CeFePO by means of angle-resolved photoemission (ARPES) in
order to understand possible reasons for the quenching of
superconductivity. We find that closely below $E_F$ both the
position and the dispersion of the valence bands are strongly
changed with respect to the ones in LaFePO what is at least partly
due to interactions with the Ce 4$f$ states. Hybridization of the
Fe 3$d$-derived VBs and the Ce 4$f$ states leads around the
$\bar\Gamma$ point of the surface BZ to strong 4$f$ admixture to
the valence bands, accompanied by a reconstruction of the Fermi
surface and a shift of the 4$f$-derived quasiparticle band to
lower binding energies.

% CeFePO single crystals were synthesized using a two step Sn-flux technique: In a first step, P and Sn were heated up to 600$^{\circ}$\,C for 5\,h using an aluminum oxide crucible which was sealed inside an evacuated silica ampule. In a second step, Ce, Fe, SnO$_2$ and Sn were added and the aluminum oxide crucible was sealed inside a Ta-container under argon atmosphere. The mixture was then heated up to 1500$^{\circ}$\,C within 5 days followed by cooling down to room temperature. The excess Sn was dissolved in hydrochloric acid.

Experiments were performed at the "$1^3$-ARPES" setup at BESSY
(Berlin) as described in Ref.~[\onlinecite{Inosov2008}], at
temperatures around 10\,K, on single crystals grown from a Sn flux
as specified in Ref.~[\onlinecite{Krellner2008}]. Due to setup
geometry, the vector potential  $\bm{A}$ of incident light is
parallel to sample surface at vertical polarization (VP) and
posses an additional perpendicular component at horizontal
polarization (HP). Dipole matrix elements for the photoexcitation
depend on the spatial extension of the orbital along the direction
of $\bm{A}$. This means that in normal emission geometry states of
$d_{3z^2-r^2}$ symmetry will contribute only at HP, while those of
$d_{xz,yz}$ and $d_{x^2-y^2}$ ($d_{xy}$, depending on the
orientation of the sample in the $(x,y)$ plane) symmetry will be
detected at both VP and HP $-$ though with different relative
intensities.

\begin{figure}[bpt]
    \includegraphics[width=8.5cm]{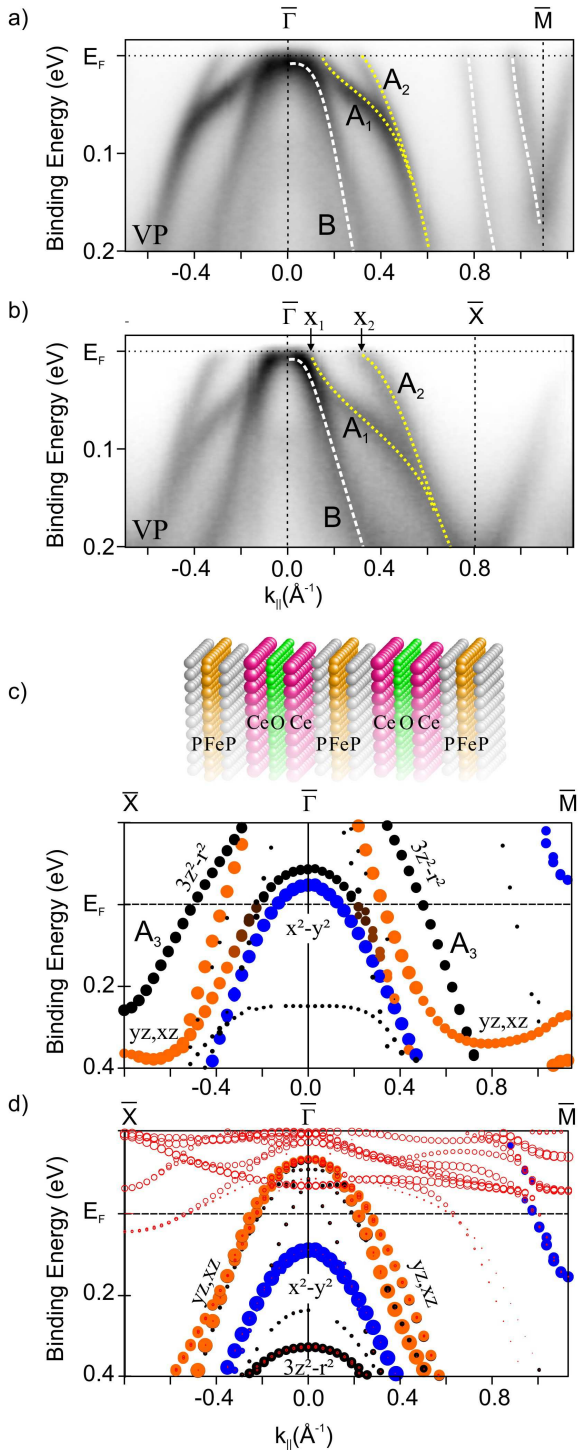}
     \caption{(Color online) Experimental ARPES images recorded from
     CeFePO at \textit{h}$\nu$=112\,eV and VP along the $\bar\Gamma$ - $\bar
     M$ (a) and $\bar\Gamma$ -$\bar X$ (b) directions in the surface
     BZ, and calculated energy bands for a slab containing
     15 atomic layers, with a P terminated surface, treating 4$f$ states
     as quasi-core (c) and valence states (d). Size of the dots indicates
     contribution of $d$ orbitals of the outermost Fe layer (solid dots)
     or of Ce 4$f$ states (4th layer, open dots). The labels indicate the
     orbitals with strongest contribution to the bands.}\label{ARPES}
\end{figure}

Photoemission (PE) spectra of Ce systems reveal a well known double-peak structure
consisting of a component at about 2\,eV BE, roughly
reflecting the 4$f^0$ final state expected from PE excitation of
an unhybridized 4$f^1$ ground state, and a feature close to $E_F$
that is only due to hybridization and reproduces the ground-state
configuration of mainly 4$f^1$ character. In our measurements we
made use of strong variations of the 4$f$ photoionization cross
section around the 4$d\rightarrow$ 4$f$ absorption threshold due
to a Fano resonance: 4$f$ emission becomes resonantly enhanced (suppressed) at
\textit{h}$\nu$=121\,eV (112\,eV) photon energy\cite{Mol1997}.

Valence-band maps taken at VP and a photon energy of 112\,eV are
shown in Fig.~\ref{ARPES}(a) and (b) for two high symmetry
directions in the surface Brillouin zone. Along the $\bar\Gamma$
-$\bar X$ direction two energy bands cross $E_F$ at x$_1\approx$
0.1 $\overline{\Gamma X}$ ($A_1$) and x$_2\approx$
0.4\,$\overline{\Gamma X}$ ($A_2$), respectively. In LaFePO
similar bands are observed but the crossings occur closer to the
$\bar X$ point at x$_1\approx$ 0.2 $\overline{\Gamma X}$ and
x$_2\approx$ 0.7 $\overline{\Gamma X}$ \cite{Lu2008}. In the
vicinity of the $\bar M$ point two additional bands can be
distinguished [Fig.~\ref{ARPES}(a), dashed], that merge in LaFePO.
All these bands are discussed  in Ref.~[\onlinecite{Lu2008}] on the
basis of LDA bulk band-structure calculations, using internally
relaxed parameters and rescaling calculated band energies by a
factor of two. In this way, the Fermi level crossings
x$_1$ and x$_2$ are caused by $d_{xz, yz}$ and
$d_{3z^2-r^2}$-derived states, respectively. The latter
should hardly be visible at VP and hence at least for the
present measurement a different character of the $A_2$ band has to
be concluded. Another parabolic hole-like band (labeled $B$) comes
very close to $E_F$ and has no direct counterpart in LaFePO.

In order to take account of the surface sensitivity of ARPES and
the fact that band positions of surface and subsurface atomic
layers may be different in BE with respect to the bulk ones
\cite{Vyalikh2009}, slab calculations were performed by means of
the linear-muffin-tin-orbital (LMTO) method \cite{And75}. It
follows from the structural and cohesive properties that the
CeFePO crystal can be cleaved mainly between the FeP and CeO
stacks, so that the surface will be terminated either by P or Ce
atoms. In the case of a P terminated slab the Fe atoms occupy the
second (subsurface) layer and the main contribution to the PE
intensity stems from the high cross section Fe 3$d$-derived bands.
A schematic view of the 15 atomic layer thick slab as well as
results of the respective slab calculations are shown in
Fig.~\ref{ARPES}(c). Note that 4$f$ states were treated as
quasi-core states in order to avoid the well known failures of LDA
in describing strongly localized states. The effect of the surface
to the observed bands can be explained to some extent with the
spatial orientation of the involved Fe $d$ states. The calculated
band structure of Fe layer in the center of the slab is very close
to the bulk band structure.

Band $B$ is quite well described by $d_{x^2-y^2}$ states which are
not strongly influenced by surface effects, since these orbitals
are oriented in the ($x$,$y$) plane and contribute to the Fe$-$P
bonds but with negligible overlap with Ce states. Two bands of
$d_{xz}$ and $d_{yz}$ symmetry cross the Fermi level in the same
way as bands $A_1$ and $A_2$. Close to the $\bar \Gamma$ point
band $A_1$ reveals increasing $d_{3z^2-r^2}$ character. Besides
these bands the calculation predicts a further band ($A_3$) of
$d_{3z^2-r^2}$ character closer to the $\bar X$ point, resembling
the situation reported in LaFePO \cite{Lu2008}. However, this band
does not appear in the above ARPES maps, because this emission is
symmetry forbidden in the case of VP excitation. Our calculations
show that the $d_{3z^2-r^2}$ states (and to a minor degree bands
of $d_{xz}$ and $d_{yz}$ symmetry) overlap with the adjacent Ce
layer where they exhibit linear combinations of $f$ symmetry at
the Ce sites, and are thus allowed for hybridization with the Ce
4$f$ states. The experimentally observed behavior of band $A_1$
may reflect effects of such hybridization, since it strongly
deviates from a parabolic dispersion. In order to get a rough
estimate of this effect, results of calculations, where the 4$f$
states are treated as valence band states, are shown in
Fig.~\ref{ARPES}(d). Due to their interaction with the $f$ states,
Fe $d_{3z^2-r^2}$ do not contribute anymore to the band structure
close to $E_F$. Instead, a band of this symmetry appears at about
0.33\,eV BE at the $\bar\Gamma$ point.

\begin{figure}[bt]
    \includegraphics[width=8.0cm]{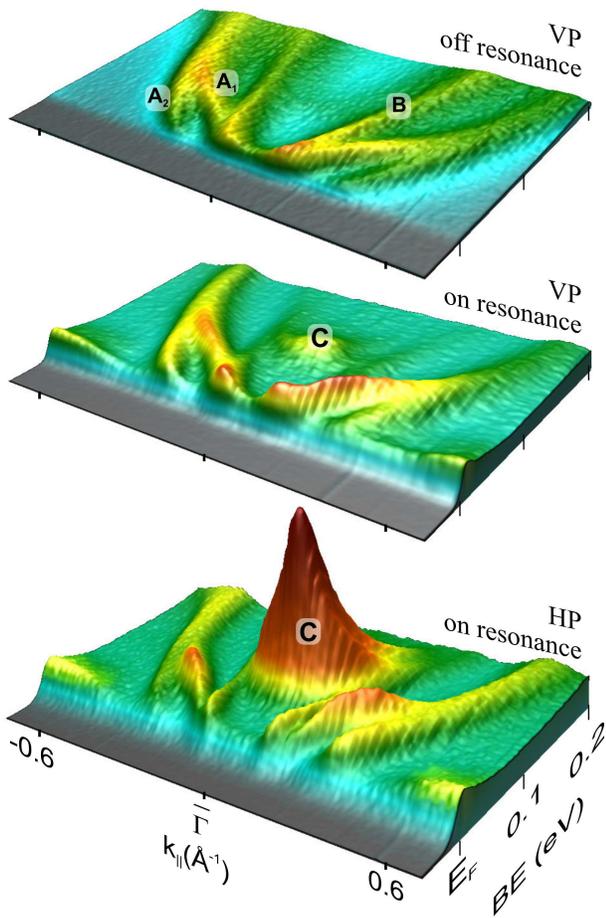}
    \caption{(Color online) ARPES images taken along the $\bar\Gamma$ - $\bar M$
    direction with the VP light at \textit{h}$\nu$=112\,eV (top, off resonance
    for $f$ emission), 121\,eV (middle, on resonance for $f$ emission), and
with the HP light
    at \textit{h}$\nu$=121\,eV (bottom, sensitive to the $d_{3z^2-r^2}$ orbitals).}\label{ResPES}
\end{figure}

An investigation of the discussed hybridization between the Fe
3$d_{3z^2-r^2}$-derived bands and the Ce 4$f$ states is possible
enhancing the cross section of photoexcitation by switching from
VP to HP (3$d$ bands) and exploiting the 4$d \rightarrow$ 4$f$
Fano resonance (4$f$ states). Respective PE maps are shown in
Fig.~\ref{ResPES}. In the topmost map, taken with VP at
\textit{h}$\nu$=112\,eV bands $A_1$, $A_2$ and $B$ are of
comparable intensity, reflecting their Fe 3$d_{xz, yz}$ and
3$d_{x^2-y^2}$ character. Switching to \textit{h}$\nu$=121\,eV the
intensity of bands $A_1$ and $A_2$ becomes essentially larger as
compared to that of band $B$. This is caused by resonant
enhancement of partial 4$f$ admixtures to the former bands.
Especially the intensity of band $A_1$ grows strongly between
0.1\,eV BE and the Fermi level, supporting the former assumption
about the hybridization with Ce 4$f$ states. In addition two other
features appear: ({\it i}) a peak directly at $E_F$ that reflects
the Ce 4$f^1$ final state and ({\it ii}) a further band with its
top at about 0.1\,eV BE, labeled $C$. Finally, at HP and \textit{h}$\nu$=121\,eV
(Fig.~\ref{ResPES} bottom), band $C$ appears extremely enhanced,  indicating
its predominant 3$d_{3z^2-r^2}$ character. Thus, its visibility at 121\,eV and VP
is only due to finite Ce 4$f$ admixtures. Band $A_3$, on the other hand, is still not observed.

In the results of our calculations [see Fig.~\ref{ARPES}(d)] the
Fe 3$d_{3z^2-r^2}$-derived band at 0.33\,eV BE corresponds to band
$C$, but the calculated band has higher BE as compared to the
experiment due to well known overestimation of the 4$f$-VB
interaction obtained with LDA. In Fig.~\ref{ARPES}(c) this band is
absent (the respective subsurface Fe 3$d$ states form band $A_3$),
however, a similar band [small dots in Fig.~\ref{ARPES} (c)] is
found at 0.25\,eV BE which is derived from the Fe 3$d$ states of
the central (bulk) layer. Thus, a possible presence of Ce at the
surface may influence band $C$ and other bands of 3$d_{3z^2-r^2}$
character.

\begin{figure}[bt]
    \includegraphics[width=8.0cm]{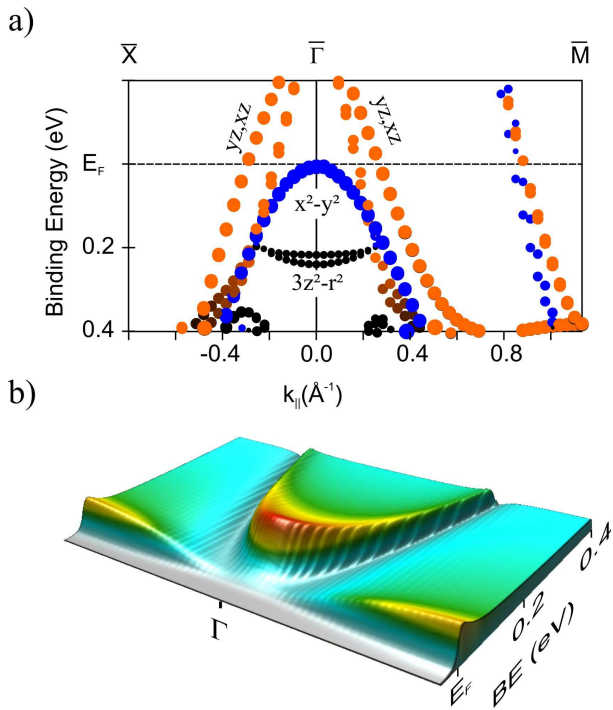}
    \caption{(Color online) (a) Calculated energy bands for a Ce
    terminated slab constructed by interchanging the FeP and CeO
    stacks of the slab shown in Fig.~\ref{ARPES}(c). The meaning
    of the symbols is the same as in Fig.~\ref{ARPES}(c) and (d).
    (b) Distribution of the spectral intensity calculated by means
    of the Periodic Anderson Model.}\label{Calc}
\end{figure}

In order to investigate this effect, the calculations where
repeated for a Ce terminated slab [see Fig.~\ref{Calc}(a)]. The
results reveal at the $\bar \Gamma$ point the formation of a
surface-derived band of 3$d_{3z^2-r^2}$ symmetry
close to the experimentally obtained position of band $C$, while
band $A_3$ is not observed. The remaining band structure looks
quite similar to the one calculated for the P terminated slab.

One can see in Fig.~\ref{ARPES}(d), that the
lowest lying Ce 4$f$-derived band is pushed above $E_F$ in that
regions of \textbf{k} space, where it interacts with the valence
bands. This is in interesting correspondence to the experimentally
observed behavior of the 4$f^1$-derived feature at $E_F$
[Fig.~\ref{ResPES}, on resonance]: Around the $\bar\Gamma$ point
this feature disappears and seems to be pushed across the Fermi
energy by the parabolic valence bands, that in turn reveal certain
4$f$ admixtures in this region of \textbf{k} space. Similar
interaction phenomena have been reported for the Yb 4$f^{13}$ bulk
emission of the heavy-fermion system YbRh$_2$Si$_2$
\cite{Vyalikh2009} as well as for the respective surface component
of YbIr$_2$Si$_2$ \cite{Danz2006, Danz2007}. In the latter case,
where the 4$f$ emission is relatively far away from the Fermi
energy (0.6\,eV BE), the phenomenon could be described
quantitatively in the light of a simplified approach to the
Periodic Anderson Model (PAM) where 4$f$ dispersion and 4$f$
admixtures to the valence bands are explained by linear
combinations of 4$f$ and valence-band states. For 4$f$ emissions
at $E_F$ the mentioned approach is, unfortunately, not applicable
because interaction with unoccupied valence states is not properly
considered. In order to solve this problem we present in the
following an elaborated approach to PAM based on dynamical
mean-field theory (DMFT).

For a numerical simulation of hybridization effects within PAM we
consider a valence band of bandwidth $W$=1.2\,eV and center at
$\epsilon_d$=0.7\,eV BE, with parabolic dispersion in the relevant
part of {\bf k} space and a 4$f$ state at $\epsilon_f$=2\,eV BE.
The self-energy was calculated by DMFT in a way as recently
proposed in Ref.~\cite{Sordi2007} but applying the noncrossing
approximation (NCA) \cite{Bickers1987} as impurity solver like in
Ref.~[\onlinecite{Lombardo2006}]. With a hybridization parameter
$t_{df}$=0.3\,eV and an on-site Coulomb repulsion $U$=7\,eV the
DMFT equation provides the self-energy of the hybridized 4$f$
states. Results in Fig.~\ref{Calc}(b) show, that the peak at $E_F$
is caused by $f$-$d$ hybridization and might be interpreted as the
tail of the Kondo resonance, which is located above
$E_F$~\cite{Reinert2001}. For those {\bf k} values where the VB
comes close to the Fermi level, the $f$ state is pushed towards
lower BE (above $E_F$) as reflected in the PE spectrum by a
decrease of 4$f$-derived intensity at $E_F$, while the intensity
of the interacting valence band becomes enhanced by substantial
4$f$ admixtures.

In our study, we compared ARPES data of CeFePO with results of LDA
slab calculations and analyzed the effect of $f$-$d$ hybridization
both in the framework of LDA and PAM. Without adjustment of
internal lattice parameters, our slab calculations reproduce
qualitatively the observed band dispersions and characters,
demonstrating the importance of surface effects in the electronic
structure. Particularly the termination of the surface either by P
or Ce atoms affects strongly shape and position of the bands. For
an interpretation of the experimental data a coexistence of both
terminations must be considered.

Strongest influence of the surface effects is found for the Fe
3$d_{3z^2-r^2}$ orbitals, which have largest overlap and,
therefore, mostly pronounced interaction with the Ce-derived
states. As a consequence, a missing Ce surface layer induces the
formation of surface-derived bands which are not observed in bulk
band structure. In LaFePO the Fe 3$d_{3z^2-r^2}$-derived states
form a pocket in the Fermi surface around the $\bar \Gamma$ point,
that is reproduced by our slab calculations if interactions with the
4$f$ states are neglected [Fig.~\ref{ARPES}(c)]. In CeFePO this
pocket is missing due to the $f$-$d$ hybridization
[Fig.~\ref{ARPES}(d)].

The Fe 3$d_{xz, yz}$-derived states are not so strongly affected
by the hybridization. Two bands of this symmetry cross the Fermi
level near $\bar \Gamma$, while two others exhibit intersections
near the $\bar M$ point. In LaFePO, each pair of these bands
nearly degenerate, forming Fermi pockets around the $\bar \Gamma$
and $\bar M$ points, respectively. The different behavior of these
bands in CeFePO might be also a consequence of the $f$-$d$
hybridization.

Superconductivity depends crucially on electronic interactions
between different FS sheets. Following the discussion in
Ref.~[\onlinecite{Kuroki2009}], it is governed by nesting between
a sheet around the $M$ point and sheets at $\Gamma$ formed by
bands of $d_{xz, yz}$ and $d_{xy}$ symmetry, respectively. Thus
the strong modifications of the Fermi surface as induced by the Ce
4$f$ states suppress superconductivity in CeFePO, which is
observed in other $R$FePO compounds without strong $f$-$d$
correlation. On the other hand, also the 4$f$ states are heavily
affected by interaction with the valence bands as reflected by the
observed dispersion of the Kondo resonance and may be important
for the understanding of quenching of magnetism and appearance of
heavy-fermion properties in CeFePO.

This work was supported by the DFG projekt VY64/1-1, and by the
Science and Technology Center in Ukraine (STCU), grant 4930. The
authors would like to thank S. Borisenko for support at "1$^3$-ARPES" beam line at BESSY.

\end{document}